\documentclass[Physsubmission, Phys]{SciPost}
\hypersetup{
    colorlinks,
    linkcolor={red!50!black},
    citecolor={blue!50!black},
    urlcolor={blue!80!black}
}

\usepackage[bitstream-charter]{mathdesign}
\DeclareSymbolFont{usualmathcal}{OMS}{cmsy}{m}{n}
\DeclareSymbolFontAlphabet{\mathcal}{usualmathcal}

\def\bkg#1{\widehat{#1}}
\def\s{\mathrm{s}}

\def\G{\Gamma}

\begin{document}

\begin{center}{\Large \textbf{
Background Field Method and Generalized Field Redefinitions\\ in Effective Field Theories\\
}}\end{center}

% TODO: write the author list here. Use initials + surname format.
% Separate subsequent authors by a comma, omit comma at the end of the list.
% Mark the corresponding author with a superscript *.
\begin{center}
Andrea Quadri\textsuperscript{1$\star$}
\end{center}

\begin{center}
{\bf 1} INFN, Sezione di Milano, via Celoria 16, I-20133 Milan, Italy\\
* andrea.quadri@mi.infn.it
\end{center}

\begin{center}
\today
\end{center}

% For convenience during refereeing (optional),
% you can turn on line numbers by uncommenting the next line:
%\linenumbers

\definecolor{palegray}{gray}{0.95}
\begin{center}
\colorbox{palegray}{
  \begin{tabular}{rr}
  \begin{minipage}{0.1\textwidth}
    \includegraphics[width=35mm]{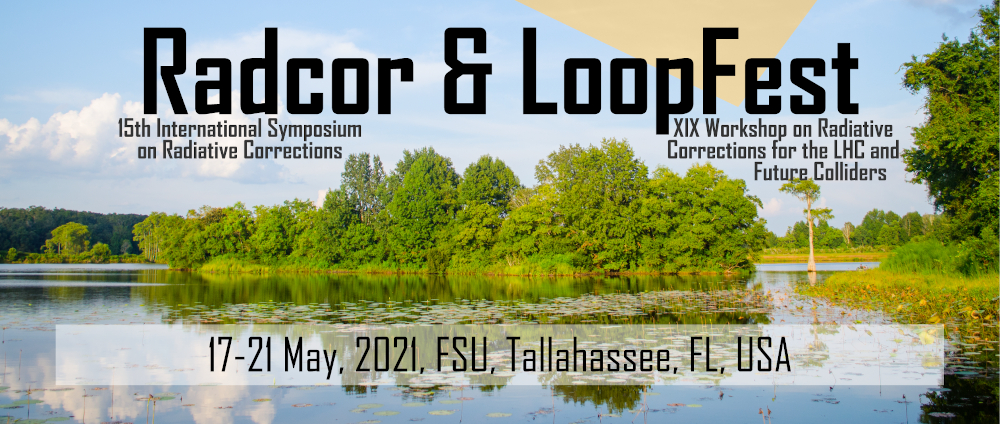}
  \end{minipage}
  &
  \begin{minipage}{0.85\textwidth}
    \begin{center}
    {\it 15th International Symposium on Radiative Corrections: \\Applications of Quantum Field Theory to Phenomenology,}\\
    {\it FSU, Tallahasse, FL, USA, 17-21 May 2021} \\
    \doi{10.21468/SciPostPhysProc.?}\\
    \end{center}
  \end{minipage}
\end{tabular}
}
\end{center}

\section*{Abstract}
{\bf
% TODO: write your abstract here.
We study the non-linear background field redefinitions arising at the quantum
level in a spontaneously broken effective gauge field theory. The non-linear
field redefinitions are crucial for the symmetric (i.e. fulfilling
all the relevant functional identities of the theory)
renormalization of gauge-invariant operators.
In a general $R_\xi$-gauge the classical background-quantum splitting is also non-linearly deformed
 by radiative corrections. In the Landau gauge 
these deformations vanish to all orders in the loop expansion.
}

\vspace{10pt}
\noindent\rule{\textwidth}{1pt}
\tableofcontents\thispagestyle{fancy}
\noindent\rule{\textwidth}{1pt}
\vspace{10pt}

\section{Introduction}
\label{sec:intro}
% TODO: write your article here.

Spontaneously broken effective gauge field theories~\cite{Buchmuller:1985jz} have become in recent years
an increasingly important
phenomenological
tool~\cite{Grzadkowski:2010es}, since they
allow to parameterize in a model-independent way possible signals
of new physics 
beyond the Standard Model (BSM) at the LHC~\cite{deFlorian:2016spz,Brivio:2017vri}.

Spontaneously broken effective gauge field theories involve
higher dimensional operators suppressed by some large mass scale $\Lambda$.
These models are not power-counting renormalizable and therefore 
they give rise to new ultra-violet (UV) divergences order by order in the loop expansion.

Nevertheless, in these models the recursive subtraction of  UV divergences
can still be achieved by local (in the sense of formal power series in the
fields, the external sources and their derivatives) counter-terms in such a way
to fulfill the so-called Batalin-Vilkovisky (BV) master equation~\cite{Gomis:1994he} (equivalently,
the Slavnov-Taylor (ST) identity~\cite{Slavnov:1972fg,Taylor:1971ff}, encoding
the BRST invariance of the gauge-fixed classical action), provided that the underlying gauge group
is non-anomalous.

This result has been established long ago in Ref.~\cite{Gomis:1995jp}
by proving that the UV divergences can be  subtracted
by a combination of generalized (i.e. non-linear) field renormalizations
and a redefinition of the coupling constants of gauge-invariant operators.

The fulfillment of the BV master equation (equivalently,
of the ST identity) ensures that physical unitarity holds,
i.e. that the cancellations of intermediate unphysical ghost
states happens~\cite{Curci:1976yb,Becchi:1974xu,Ferrari:2004pd}.

Due to the complexity of these computations, attempts have been made
to implement  the background field method (BFM)~\cite{DeWitt:1967ub,KlubergStern:1974xv,Abbott:1980hw,Abbott:1983zw,Denner:1994xt,Grassi:1999nb,Grassi:1995wr,Ferrari:2000yp} technique in order to simplify the evaluation of the required counter-terms
in the context of the (geometric) Standard Model Effective Field
Theory (SMEFT)~\cite{Helset:2018fgq,Corbett:2019cwl,Corbett:2020ymv}.

The main virtue of the BFM is that it allows to retain (background) gauge invariance to
all orders in perturbation theory. The resulting
background Ward identity is linear in the quantum
fields, unlike the ST identity, and hence the additional relations imposed by the BFM on the one-particle irreducible (1-PI) Green's functions are easier to study. 

For instance, in the power-counting renormalizable case,
with linear field and background redefinitions, the BFM 
ties together the background gauge field $\bkg{\Phi}$ and the 
coupling constant renormalizations so that  the charge renormalization factor can be obtained by evaluating just the gauge two-point 1-PI amplitude.

An important question is to clarify whether the same simplifications
arise in the context of effective field theories. 
In fact, while the background Ward
identity is still linear, the field and background field redefinitions
are in general not~\cite{Anselmi:2015niw,Quadri:2021syf}.
More precisely one finds that that~\cite{Quadri:2021syf}:
{\em i)} the tree-level background-quantum
   splitting
   $$ \Phi = \bkg{\Phi} + Q_\Phi $$
   is in general deformed in a non-linear
    (and gauge-dependent) way, unlike in the power-counting
   renormalizable case where only 
   multiplicative $Z$-factors arise
   both for
   background and quantum fields;
{\em ii)} in the Landau gauge no deformation of the tree-level
   background-quantum splitting happens, to all orders in the loop expansion;
{\em iii)} also the 
   background fields renormalize non-linearly, similarly to what happens to  quantum fields when no backgrounds are switched on~\cite{Binosi:2020unh,Binosi:2019olm,Binosi:2019nwz};
{\em iv)}  the redefinition of the background fields is background gauge invariant. This follows from non-trivial
   cancellations between the 
   non gauge-invariant contributions to
   the background-quantum splitting
    and the
   non gauge-invariant terms in the generalized field redefinitions of the quantum fields;
%    \item a noticeable exception
%    is the background Landau gauge,
%    in which the background field redefinitions are background gauge-invariant;
{\em v)} both the 
    background and the quantum field
    redefinitions need to be taken into account in order
    to ensure gauge-invariance of the
    coupling constants renormalization.

Full control of the background and quantum fields renormalization
is required in order to consistently subtract the SMEFT at higher orders in the loop expansion. 

For the sake of simplicity 
 we study  here the Abelian Higgs-Kibble model
supplemented by dim.6 operators. This is a toy model illustrating the new features of the BFM arising in the
effective field theory context. The techniques and the results presented here  can be easily generalized to
non-Abelian gauge groups.

\section{The model and its symmetries}

We consider the Abelian Higgs-Kibble model supplemented by 
the following set of dimension $6$ operators:
{\small
\begin{align}
    \int  \,  \Bigg [ \frac{z}{2v^2} \partial^\mu ( \phi^\dagger \phi) \partial_\mu ( \phi^\dagger \phi) + \frac{g_1}{\Lambda^2} 
    \Big ( \phi^\dagger \phi - \frac{v^2}{2} \Big )
    (D^\mu \phi)^\dagger (D_\mu \phi) %%%
	+  \frac{g_2}{\Lambda^2} \Big ( \phi^\dagger \phi - \frac{v^2}{2} \Big ) F_{\mu\nu}^2 
	+  \frac{g_3 }{6 \Lambda^2} \Big ( \phi^\dagger \phi - \frac{v^2}{2} \Big )^3 \Bigg ] \, .
\end{align}
}
We work in the so-called $X$-formalism~\cite{Quadri:2006hr,Quadri:2016wwl,Binosi:2017ubk}. In the latter 
 an additional set of auxiliary fields $X_{1,2}$ is introduced, $X_1$ being
a Nakanishi-Lautrup field.
The gauge-invariant field coordinate $X_2$, that
describes the physical scalar mode, obeys the following on-shell condition (imposed by the $X_1$-equation of motion):
\begin{align}
    X_2 \sim \frac{1}{v} \Big ( \phi^\dagger \phi  - \frac{v^2}{2} \Big ) \, .
    \label{X2.def}
\end{align}
The convenience of the $X$-formalism is related to the
existence of a set of functional identities fixing the 
1-PI amplitudes of the auxiliary fields in terms of ancestor amplitudes with a better UV degree of divergence
than in the ordinary formalism.
One recovers the results for the model in the standard approach by going on-shell with $X_{1,2}$. The procedure has been studied to all orders in the loop expansion in~Refs.\cite{Quadri:2006hr,Quadri:2016wwl,Binosi:2017ubk}.

We adopt the following background gauge-fixing condition:
\begin{align}
    \bkg{\cal F}_\xi = \partial^\mu (A_\mu - \bkg{A}_\mu) + ~ \xi e (\bkg{\phi}_0 \chi - \bkg{\chi}\phi_0) \, , 
    \label{bkg.g.f}
\end{align}
Hats denote the background fields, $A_\mu$ is the Abelian gauge field and $\phi \equiv \frac{1}{\sqrt{2}} (\phi_0 + i \chi)$ is the scalar doublet, with  $\phi_0 = \sigma + v$.  Notice that the replacement in Eq.(\ref{X2.def}) is carried out in the classical action
only whenever the combination $\phi^\dagger \phi - \frac{v^2}{2}$ appears. The complete classical
vertex functional of the theory is reported in Ref.~\cite{Quadri:2021syf}.

The gauge-fixed model exhibits a U(1) gauge BRST invariance, obtained by promoting the infinitesimal parameter of the gauge transformation of the fields to a ghost $\omega$ and then 
supplementing the transformation of the antighost $\bar \omega$ into the Nakanishi-Lautrup field $b$:
\begin{align}
& s A_\mu = \partial_\mu \omega \, ; \quad s\phi = i e \omega \phi \,; \quad s \omega = 0 \, ; \quad s \bar \omega = b \, ; \quad s b= 0 \, . 
\end{align}
A further BRST symmetry encodes the decoupling  of the $X_1$-modes 
from physical states  and is given by ($c, \bar c$ are the ghost and antighost associated with this additional
{\em constraint} BRST invariance):
\begin{align}
	\s X_1 = v c; \, \quad \s c  = 0; \quad \s \bar c = \phi^\dagger\phi - \frac{v^2}{2} - v X_2 \, .
\end{align}

The quantum dependence  of the effective action 
on the background fields can be constrained by algebraic tools based on the extension of the BRST differential in such a way that for each background field a classical ghost partner is introduced, i.e. 
$$
s \bkg{\Phi} = \Omega_{\bkg{\Phi}} \, ,
    \qquad s \Omega_{\bkg{\Phi}}  = 0 \, .
$$
The vertex functional $\G$ obeys then the following
Slavnov-Taylor (ST) identity:
\begin{align}
	& {\cal S}(\G)  = \int \mathrm{d}^4x \, \Big [ 
	\partial_\mu \omega \frac{\delta \G}{\delta A_\mu} + \frac{\delta \G}{\delta \sigma^*} \frac{\delta \G}{\delta \sigma}  + \frac{\delta \G}{\delta \chi^*} \frac{\delta \G}{\delta \chi} 
	+ b \frac{\delta \G}{\delta \bar \omega} 
	+ \Omega_\mu \frac{\delta \G}{\delta \bkg{A}_\mu} +
	\Omega_{\bkg{\sigma}} \frac{\delta \G}{\delta \bkg{\sigma}} + \Omega_{\bkg{\chi}} \frac{\delta \G}{\delta \bkg{\chi}}
	\Big ] = 0 \,
	\label{sti} 
	\end{align}
and in addition the following linear background Ward identity:
\begin{align}
   {\cal W} (\Gamma ) = &
      - \partial^\mu \frac{\delta \G}{\delta A^\mu}
    - e \chi \frac{\delta \G}{\delta \sigma} 
    + e (\sigma + v) \frac{\delta \G}{\delta \chi} %\nonumber \\
    %& 
      - \partial^\mu \frac{\delta \G}{\delta \bkg{A}^\mu}
    - e \bkg{\chi} \frac{\delta \G}{\delta \bkg{\sigma}} 
    + e (\bkg{\sigma} + v) \frac{\delta \G}{\delta \bkg{\chi}} \nonumber \\
    & 
     - e \chi^* \frac{\delta \G}{\delta \bkg{\sigma^*}} 
    + e \sigma^* \frac{\delta \G}{\delta \chi^*} = 0 \, .
    \label{bkg.wi}
\end{align}
In the above equations $\sigma^*, \chi^*$ denote the 
external sources coupled to the non-linear BRST transformations of the fields (the so-called antifields~\cite{Gomis:1994he}).

\section{Quantum background-splitting deformation}
\label{sec:bkg-q-split}

By taking a derivative of the ST identity in Eq.(\ref{sti}) w.r.t $\Omega_\mu, \Omega_{\bkg{\sigma}}, \Omega_{\bkg{\chi}}$ and then 
setting all the fields and external sources with positive ghost number to zero as well as $b=0$ we get~\footnote{We denote by a subscript the functional differentiation w.r.t. the argument, i.e. $\G_X = \frac{\delta \G}{\delta X}$.}
\begin{align}
\G'_{\bkg{\sigma}} = - \int 
\Big [
\G'_{\Omega_{\bkg{\sigma}} \sigma^*} \G'_{\sigma} +
\G'_{\Omega_{\bkg{\sigma}} \chi^*} \G'_{\chi} \Big ] \, , \qquad \G'_{\bkg{\chi}} = - \int 
\Big [
\G'_{\Omega_{\bkg{\chi}} \sigma^*} \G'_{\sigma} \G'_{\Omega_{\bkg{\chi}} \chi^*} \G'_{\chi} \Big ] \, .
\label{bkg.eoms}
\end{align}
In the above equation we have denoted by a prime the 
functionals evaluated at zero ghost number and at $b=0$.
Let us project Eq.(\ref{bkg.eoms}) 
at first order in the loop expansion.

There is no tree-level 1-PI amplitude involving
the antifields $\sigma^*,\chi^*$
together with the background ghosts
$\Omega_{\bkg{\sigma}},\Omega_{\bkg{\chi}}$,
since the background field only enters in the gauge-fixing.\footnote{A different basis is often equivalently used in BFM calculations, namely the quantum fields are defined as $Q_\mu=A_\mu - \bkg{A}_\mu,q_\sigma = \sigma - \bkg{\sigma}, q_\chi = \chi- \bkg{\chi}$. We prefer the basis
($A_\mu, \sigma, \chi$) since it simplifies the solution of the
extended ST identity in Eq.(\ref{bkg.eoms}), due to the fact that
the dependence on the background fields in limited to the
gauge-fixing sector}
Thus
\begin{align}
\G^{(1)'}_{\bkg{\sigma}} = - \int 
\Big [
\G^{(1)'}_{\Omega_{\bkg{\sigma}} \sigma^*} \G^{(0)'}_{\sigma} +
\G^{(1)'}_{\Omega_{\bkg{\sigma}} \chi^*} \G^{(0)'}_{\chi} \Big ] \, , \qquad
& \G^{(1)'}_{\bkg{\chi}} = - \int 
\Big [
\G^{(1)'}_{\Omega_{\bkg{\chi}} \sigma^*} \G^{(0)'}_{\sigma} +
\G^{(1)'}_{\Omega_{\bkg{\chi}} \chi^*} \G^{(0)'}_{\chi} \Big ] \, .
\label{bkg.eoms.1loop}
\end{align}
Since we look at the UV divergent part of the amplitudes,
we can limit ourselves to solve the above equation
for local functionals. Again at $b=0$ there is no background
dependence of the classical action.
Eq.(\ref{bkg.eoms.1loop}) fixes the full dependence on the background fields in terms of the amplitudes at zero background, once the
kernels $\G^{(1)'}_{\Omega_{\bkg{\sigma}} \sigma^*},\G^{(1)'}_{\Omega_{\bkg{\sigma}} \chi^*},\G^{(1)'}_{\Omega_{\bkg{\chi}} \sigma^*}, \G^{(1)'}_{\Omega_{\bkg{\chi}} \chi^*}$
are known.
The kernels  are gauge-dependent. 
Therefore we consider the
Feynman gauge and the Landau gauge separately.

\subsection{Feynman gauge}

In the Feynman gauge $\xi=1$ the kernels are non-vanishing.
%and therefore we 
%must take into account their %contribution
%in order to recover the background %dependence of the vertex functional.
%For that purpose 
By power-counting they  contain at most logarithmic divergences.
The kernels have been computed to the accuracy required to renormalize
dim.6 operators in Ref.~\cite{Binosi:2020unh}.
Let us denote by a bar the UV divergent part of a given amplitude or of the full vertex functional.

We notice that the kernels in the $(A_\mu, \sigma, \chi)$-basis
do not depend on the background fields,
so by integrating Eq.(\ref{bkg.eoms.1loop}) 
and projecting on the UV divergent sector
we find
a linear dependence on the $\bkg{\Phi}$ 's:
\begin{align}
 \overline{\G}^{(1)'} = - \int  \Big [
 \Big ( \bkg{\sigma} ~ \overline{\G}^{(1)'}_{\Omega_{\bkg{\sigma}} \sigma^*} + \bkg{\chi} ~
 \overline{\G}^{(1)'}_{\Omega_{\bkg{\chi}} \sigma^*} 
 \Big ) \G^{(0)'}_{\sigma} + 
 \Big ( \bkg{\sigma} ~ \overline{\G}^{(1)'}_{\Omega_{\bkg{\sigma}} \chi^*} + \bkg{\chi}~ \overline{\G}^{(1)'}_{\Omega_{\bkg{\chi}} \chi^*} 
 \Big ) \G^{(0)'}_{\chi} \Big ] + 
 \left . \overline{\G}^{(1)'} \right |_{\bkg{\sigma} = \bkg{\chi} = 0} \, .
 \label{bkg.1loop.feyn}
\end{align}
%

%
%We notice that the r.h.s. of the above equation
%turns out to be linear in the %background fields,
%as a consequence of the vanishing of %the $\gamma$'s coefficients
%involving at least one background leg.

The last term in Eq.(\ref{bkg.1loop.feyn})  
denotes the UV divergent part of the
vertex functional at zero background. 
It has been evaluated in~\cite{Binosi:2020unh} 
for the relevant sector of operators up to dimension $6$.
Eq.(\ref{bkg.1loop.feyn}) can be interpreted as follows.
It entails that the classical background-quantum splitting 
$\Phi = Q_\Phi + \bkg{\Phi}$ does not survive at the quantum level,
being deformed according to
\begin{align}
    & \sigma %= \bkg{\sigma} + 
    %q_\sigma 
    \rightarrow \bkg{\sigma} + q_\sigma 
      -  \bkg{\sigma} ~ \overline{\G}^{(1)'}_{\Omega_{\bkg{\sigma}} \sigma^*}- \bkg{\chi} ~
 \overline{\G}^{(1)'}_{\Omega_{\bkg{\chi}} \sigma^*}  \, , \qquad
    \chi %= \bkg{\chi} + q_\chi 
    \rightarrow \bkg{\chi} + q_\chi -
    \bkg{\sigma} ~ \overline{\G}^{(1)'}_{\Omega_{\bkg{\sigma}} \chi^*} - \bkg{\chi}~ \overline{\G}^{(1)'}_{\Omega_{\bkg{\chi}} \chi^*}  \, .
    \label{bkg.qnt.split}
\end{align}
By inserting the redefinitions in Eq.(\ref{bkg.qnt.split}) into  the tree-level vertex
functional $\G^{(0)}$, one gets back the terms in square brackets in
 Eq.(\ref{bkg.1loop.feyn}).
The kernels $\overline{\G}^{(1)'}_{\Omega_\Phi \Phi}$ 
 depend on the fields and external sources
in a complicated way. Eq.(\ref{bkg.qnt.split}) is 
therefore a highly non-linear redefinition w.r.t. the quantum fields.

\subsection{Landau gauge}

In the Landau gauge $\xi=0$ all  
the four kernels are identically zero since there are no interaction vertices that could generate them. Therefore the classical background-quantum splitting
$\Phi = \bkg{\Phi} + Q_\Phi$ does not receive any radiative corrections.
This is an all-order result.

\section{Generalized background field redefinitions}

The final form of the generalized background field redefinitions
in the target theory can be read off by going on-shell with $X_{1,2}$
through the substitution on the external sources described in
Refs.~\cite{Quadri:2021syf}. The resulting expressions are too long to be reported here and can be found in \cite{Quadri:2021syf}.
In particular at zero quantum fields $Q_\Phi=0$ they reduce (to the accuracy required for the renormalization of dim.6 operators) to the following expressions:
\begin{align}
    \begin{pmatrix}
        \bkg{\sigma}_R  \\
        \bkg{\chi}_R 
    \end{pmatrix}
= \Big [
a_0 +   
     \Big (  \frac{a_1 g_1}{\Lambda^2} + a_2 \Big ) \Big ( \bkg{\phi}^\dagger \bkg{\phi} - \frac{v^2}{2} \Big ) +
    a_3 \Big (\bkg{\phi}^\dagger \bkg{\phi} - \frac{v^2}{2} \Big )^2 + \dots
\Big ]    
\begin{pmatrix}
        \bkg{\sigma} + v \\
        \bkg{\chi}
    \end{pmatrix} \, .
\label{bgfrs.target}
\end{align}
with (gauge-dependent) coefficients $a_j$'s. Their explicit
expression has been given in Ref.~\cite{Quadri:2021syf}.
As can be seen from Eq.(\ref{bgfrs.target})
this is non-linear gauge-invariant (albeit gauge-dependent) field redefinition.

\section{Conclusion}

Application of the BFM to spontaneously broken gauge effective field theories exhibits novel features with respect to the usual
power-counting renormalizable case. In particular the background-quantum splitting is non linearly deformed by quantum corrections and also the background fields are non-trivially redefined, at variance with the power-counting renormalizable case, where both the quantum and the background fields renormalize linearly.
These redefinitions are crucial in order to preserve the locality
of the recursive subtraction of the UV divergences in a way compatible
with the symmetries of the model. The tools described here provide a constructive approach to the renormalization of the sponteaneously broken effective gauge theories in the modern sense of Ref.~\cite{Gomis:1995jp}.

\section*{Acknowledgements}
Useful discussions with D.~Anselmi and D.~Binosi are gratefully acknowledged.

% TODO: include author contributions

% TODO: include funding information
\paragraph{Funding information}
Partial financial support by INFN is acknowledged.

% TODO:
% Provide your bibliography here. You have two options:

% FIRST OPTION - write your entries here directly, following the example below, including Author(s), Title, Journal Ref. with year in parentheses at the end, followed by the DOI number.
%\begin{thebibliography}{99}
%\bibitem{1931_Bethe_ZP_71} H. A. Bethe, {\it Zur Theorie der Metalle. i. Eigenwerte und Eigenfunktionen der linearen Atomkette}, Zeit. f{\"u}r Phys. {\bf 71}, 205 (1931), \doi{10.1007\%2FBF01341708}.
%\bibitem{arXiv:1108.2700} P. Ginsparg, {\it It was twenty years ago today... }, \url{http://arxiv.org/abs/1108.2700}.
%\end{thebibliography}

% \bibliographystyle{SciPost_bibstyle} % Include this style file here only if you are not using our template
%\bibliography{bibliography_1loop.bib}

\nolinenumbers

\end{document}